\documentstyle[12pt,graphicx,amsmath]{article}
\newlength{\pubnumber} 

\catcode`\@=11
\@addtoreset{equation}{section}

\def\section{\@startsection{section}{1}{\z@}{3.5ex plus 1ex minus .2ex}
 {2.3ex plus .2ex}{\large\bf}}
\def\subsection{\@startsection{subsection}{2}{\z@}{2.3ex plus .2ex}
 {2.3ex plus .2ex}{\bf}}

\newcommand{\ba}{\begin{eqnarray}}
\newcommand{\ea}{\end{eqnarray}}
\begin{document}

\begin{titlepage}
\samepage{
\setcounter{page}{1}
\rightline{LPTENS--09/28}
\rightline{LTH--843}
\rightline{October 2009}

\vfill
\begin{center}
 {\Large \bf
Exophobic Quasi--Realistic Heterotic String Vacua
}
\vspace{1cm}
\vfill {\large
Benjamin Assel$^{1}$,
Kyriakos Christodoulides$^{2}$\\
\vspace{.1in}
Alon E. Faraggi$^{2}$,
Costas Kounnas$^{3}$\footnote{Unit\'e Mixte de Recherche
(UMR 8549) du CNRS et de l'ENS
 associ\'e¿e \`a l'universit\'e¿ Pierre et Marie Curie}
 and
John Rizos$^{4}$}\\
\vspace{1cm}
{\it $^{1}$ Centre de Physique Th\'eorique,
             Ecole Polytechnique,
         F--91128 Palaiseau, France\\}
\vspace{.05in}
{\it $^{2}$ Dept.\ of Mathematical Sciences,
             University of Liverpool,
         Liverpool L69 7ZL, UK\\}
\vspace{.05in}
{\it $^{3}$ Lab.\ Physique Th\'eorique,
Ecole Normale Sup\'erieure, F--75231 Paris 05, France\\}
\vspace{.05in}
{\it $^{4}$ Department of Physics,
              University of Ioannina, GR45110 Ioannina, Greece\\}
\vspace{.025in}
\end{center}
\vfill
\begin{abstract}

We demonstrate the existence of heterotic--string vacua that are
free of massless exotic fields. The need to break the
non--Abelian GUT symmetries in $k=1$ heterotic--string models
by Wilson lines, while preserving the GUT embedding of the
weak--hypercharge and the GUT prediction
$\sin^2\theta_w(M_{\rm GUT})=3/8$, necessarily implies that the models
contain states with fractional electric charge. Such states are
severely restricted by observations, and must be confined or
sufficiently massive and diluted. We construct
the first quasi--realistic heterotic--string models in which the
exotic states do not appear in the massless spectrum, and only exist,
as they must, in the massive spectrum. The $SO(10)$ GUT symmetry
is broken to the Pati--Salam subgroup. Our PS heterotic--string
models contain adequate Higgs representations to break the GUT and electroweak
symmetry, as well as colour Higgs triplets that can be used for the missing
partner mechanism. By statistically sampling the space of Pati--Salam
vacua we demonstrate the abundance of quasi--realistic
three generation models
that are completely free of massless exotics, rendering it plausible that
obtaining realistic Yukawa couplings may be possible in this space of models.

\noindent

\end{abstract}
\smallskip}
\end{titlepage}

\setcounter{footnote}{0}

\def\beq{\begin{equation}}
\def\eeq{\end{equation}}
\def\beqn{\begin{eqnarray}}
\def\eeqn{\end{eqnarray}}

\def\no{\noindent }
\def\nolabel{\nonumber }
\def\ie{{\it i.e.}}
\def\eg{{\it e.g.}}
\def\half{{\textstyle{1\over 2}}}
\def\third{{\textstyle {1\over3}}}
\def\quarter{{\textstyle {1\over4}}}
\def\sixth{{\textstyle {1\over6}}}
\def\m{{\tt -}}
\def\p{{\tt +}}

\def\Tr{{\rm Tr}\, }
\def\tr{{\rm tr}\, }

\def\slash#1{#1\hskip-6pt/\hskip6pt}
\def\slk{\slash{k}}
\def\GeV{\,{\rm GeV}}
\def\TeV{\,{\rm TeV}}
\def\y{\,{\rm y}}
\def\SM{Standard--Model }
\def\SUSY{supersymmetry }
\def\SSSM{supersymmetric standard model}
\def\vev#1{\left\langle #1\right\rangle}
\def\l{\langle}
\def\r{\rangle}
\def\o#1{\frac{1}{#1}}

\def\Htw{{\tilde H}}
\def\chibar{{\overline{\chi}}}
\def\qbar{{\overline{q}}}
\def\ibar{{\overline{\imath}}}
\def\jbar{{\overline{\jmath}}}
\def\Hbar{{\overline{H}}}
\def\Qbar{{\overline{Q}}}
\def\abar{{\overline{a}}}
\def\alphabar{{\overline{\alpha}}}
\def\betabar{{\overline{\beta}}}
\def\tautwo{{ \tau_2 }}
\def\thetatwo{{ \vartheta_2 }}
\def\thetathree{{ \vartheta_3 }}
\def\thetafour{{ \vartheta_4 }}
\def\ttwo{{\vartheta_2}}
\def\tthree{{\vartheta_3}}
\def\tfour{{\vartheta_4}}
\def\ti{{\vartheta_i}}
\def\tj{{\vartheta_j}}
\def\tk{{\vartheta_k}}
\def\calF{{\cal F}}
\def\smallmatrix#1#2#3#4{{ {{#1}~{#2}\choose{#3}~{#4}} }}
\def\ab{{\alpha\beta}}
\def\Minv{{ (M^{-1}_\ab)_{ij} }}
\def\bone{{\bf 1}}
\def\ii{{(i)}}
\def\V{{\bf V}}
\def\N{{\bf N}}

\def\b{{\bf b}}
\def\S{{\bf S}}
\def\X{{\bf X}}
\def\I{{\bf I}}
\def\mb{{\mathbf b}}
\def\mS{{\mathbf S}}
\def\mX{{\mathbf X}}
\def\mI{{\mathbf I}}
\def\balpha{{\mathbf \alpha}}
\def\bbeta{{\mathbf \beta}}
\def\bgamma{{\mathbf \gamma}}
\def\bxi{{\mathbf \xi}}

\def\t#1#2{{ \Theta\left\lbrack \matrix{ {#1}\cr {#2}\cr }\right\rbrack }}
\def\C#1#2{{ C\left\lbrack \matrix{ {#1}\cr {#2}\cr }\right\rbrack }}
\def\tp#1#2{{ \Theta'\left\lbrack \matrix{ {#1}\cr {#2}\cr }\right\rbrack }}
\def\tpp#1#2{{ \Theta''\left\lbrack \matrix{ {#1}\cr {#2}\cr }\right\rbrack }}
\def\l{\langle}
\def\r{\rangle}
\newcommand{\cc}[2]{c{#1\atopwithdelims[]#2}}
\newcommand{\nn}{\nonumber}


\def\inbar{\,\vrule height1.5ex width.4pt depth0pt}

\def\IC{\relax\hbox{$\inbar\kern-.3em{\rm C}$}}
\def\IQ{\relax\hbox{$\inbar\kern-.3em{\rm Q}$}}
\def\IR{\relax{\rm I\kern-.18em R}}
 \font\cmss=cmss10 \font\cmsss=cmss10 at 7pt
\def\IZ{\relax\ifmmode\mathchoice
 {\hbox{\cmss Z\kern-.4em Z}}{\hbox{\cmss Z\kern-.4em Z}}
 {\lower.9pt\hbox{\cmsss Z\kern-.4em Z}}
 {\lower1.2pt\hbox{\cmsss Z\kern-.4em Z}}\else{\cmss Z\kern-.4em Z}\fi}

\def\AEF{A.E. Faraggi}
\def\JHEP#1#2#3{{\it JHEP}\/ {\bf #1} (#2) #3}
\def\NPB#1#2#3{{\it Nucl.\ Phys.}\/ {\bf B#1} (#2) #3}
\def\PLB#1#2#3{{\it Phys.\ Lett.}\/ {\bf B#1} (#2) #3}
\def\PRD#1#2#3{{\it Phys.\ Rev.}\/ {\bf D#1} (#2) #3}
\def\PRL#1#2#3{{\it Phys.\ Rev.\ Lett.}\/ {\bf #1} (#2) #3}
\def\PRT#1#2#3{{\it Phys.\ Rep.}\/ {\bf#1} (#2) #3}
\def\MODA#1#2#3{{\it Mod.\ Phys.\ Lett.}\/ {\bf A#1} (#2) #3}
\def\IJMP#1#2#3{{\it Int.\ J.\ Mod.\ Phys.}\/ {\bf A#1} (#2) #3}
\def\nuvc#1#2#3{{\it Nuovo Cimento}\/ {\bf #1A} (#2) #3}
\def\RPP#1#2#3{{\it Rept.\ Prog.\ Phys.}\/ {\bf #1} (#2) #3}
\def\EJP#1#2#3{{\it Eur.\ Phys.\ Jour.}\/ {\bf C#1} (#2) #3}
\def\etal{{\it et al\/}}

\hyphenation{su-per-sym-met-ric non-su-per-sym-met-ric}
\hyphenation{space-time-super-sym-met-ric}
\hyphenation{mod-u-lar mod-u-lar--in-var-i-ant}


\setcounter{footnote}{0}
\section{Introduction}

String theory provides a perturbative
calculational framework for quantum gravity,
while simultaneously giving rise to the gauge and matter structures
that are observed in particle experiments. Furthermore,
the gauge and matter sectors are required by its
self--consistency constraints.
Additionally, by producing spinorial matter representations
in the perturbative spectrum,
the heterotic--string naturally accommodates the $SO(10)$ embedding
of the Standard Model, which is well motivated by the experimental data.
Absence of higher order Higgs representations in heterotic--string models
that are based on level one Kac--Moody current algebras necessitates that the $SO(10)$
symmetry is broken directly at the string level by discrete Wilson lines.

A well known theorem due to Schellekens \cite{schellekens} states that
any such string model that preserves the canonical $SU(5)$-- or $SO(10)$--GUT embedding
of the weak hypercharge, and in which the non--Abelian GUT symmetries are
broken by discrete Wilson lines, necessarily contain states that carry
charges that do not obey the original GUT quantization rule \cite{schellekens}\footnote{
A similar observation was made in the context of Calabi--Yau compactification
models with $E_6$ gauge group broken by Wilson lines \cite{ww}.}.
In terms of the Standard Model charges these exotic states carry
fractional electric charge. The existence of such states is severely
constrained by experiments. Electric charge conservation implies that the
lightest of these states is stable.

Many experimental searches for fractionally charged states have been conducted \cite{halyo}.
However, no reported observation of any such particle has ever been confirmed and there
are strong upper bounds on their abundance \cite{halyo}.
This implies that such exotic states in string
models should be either confined into integrally charged states \cite{revamp}, or be
sufficiently heavy and diluted in the cosmological evolution of the universe \cite{fc}.
The first of these solutions is, however, problematic, due to the effect of the
charged states on the renormalisation group running of the weak--hypercharge
and gauge coupling unification. The preferred solution is therefore for
the fractionally charged states to become sufficiently massive, {\it i.e.}
with a mass which is larger than the GUT scale. In this case the fractionally
charged states can be diluted by the inflationary evolution of the universe.
Due to their heavy mass they will not be reproduced during re--heating and
the experimental constraints can be evaded.

By producing the gauge and matter structures, that are the building blocks of the
Standard Particle Model, string theory enables the development of a phenomenological
approach. A pivotal task in this regard, is the construction of quasi--realistic
three generation models. Such models can in turn be used to explore the properties
of string theory and quantum gravity. Quasi--realistic perturbative heterotic--string
models have been constructed \cite{revamp, fny, alr, eu, cfn, lrs, raby}
by using free fermionic \cite{fff} and orbifold
\cite{orbifold} techniques.
The existence of fractionally charged states in string models is endemic.
To our knowledge all string models constructed to date contain such states
in the massless spectrum.
One obvious remedy is to modify the GUT embedding of the weak-hypercharge,
that produces integral charges for all states. However, the result is that the
GUT prediction of $\sin^2\theta_W(M_{GUT})$ is modified, and the GUT
embedding of the Standard Model spectrum is lost.
Maintaining the GUT embedding and the canonical GUT prediction
$\sin^2\theta_W(M_{GUT})=3/8$
therefore necessitates the existence of fractionally charged
states in the physical spectrum.

While in some models the exotic states are chiral and therefore
necessarily remain massless, there exist an abundance of models in which they
appear in vector--like representations and therefore can gain mass in the
effective low energy field theory, by the Vacuum Expectation Value (VEV)
of a Standard Model singlet field. Such singlet field VEVs which are
of the order of the string scale, are in fact often mandated in the models
due to the existence of an anomalous $U(1)$, which is broken by the Green--Schwarz
mechanism. The details are model dependent, but there
exist models in which all the exotic states couple to $SO(10)$ singlet fields at the
cubic level of the superpotential \cite{fny,fc}. Assigning VEVs of order that exceeds the
GUT scale to these set of fields gives sufficiently heavy mass to the exotic states.
Indeed supersymmetric preserving solution with VEVs to the required set of fields
were found in ref. \cite{cfn}. The caveat is that this demonstration is achieved
by an effective field theory analysis. The question therefore remains whether there exist
string models that are free of massless fractionally charged states. In such models
states with fractional charge necessarily appear in the massive spectrum,
but do not appear as massless states.

In this paper we show that such string models do in fact exist. We use the free fermionic
formalism for the analysis. In the orbifold language the free fermionic construction
correspond to symmetric, asymmetric or freely acting orbifolds.
A subclass of them correspond to symmetric $Z_2\times Z_2$ orbifold
compactifications at enhanced symmetry points in the toroidal moduli
space \cite{z2z21,z2z22}. The chiral matter spectrum arises from
twisted sectors and thus does not depend on the moduli. This
facilitates the complete classification of the topological sectors of the
$Z_2\times Z_2$ symmetric orbifolds. For type II string $N=2$ supersymmetric vacua
the general free fermionic classification techniques were developed in ref.
\cite{gkr}. The method was extended in refs.
\cite{fknr,fkr} for the classification of heterotic $Z_2\times Z_2$ free
fermionic orbifolds, with unbroken $SO(10)$ and $E_6$ GUT symmetries.

Absence of adjoint Higgs representations in heterotic--string models with unbroken GUT
symmetries realised as level one Kac--Moody algebras means that the models
classified in \cite{fknr,fkr}
are not realistic. The GUT gauge symmetry must be broken directly at the
string level. In the free fermionic models the GUT gauge symmetry generated by untwisted
vector bosons is $SO(10)$, and can be enhanced to a larger gauge group by gauge
bosons arising from other sectors. Phenomenologically the most appealing case is that
of $SO(10)$
by itself, and therefore it is reasonable to demand that gauge bosons which enhance the
$SO(10)$ symmetry be projected out by the Generalised GSO (GGSO) projections.
The $SO(10)$ symmetry must therefore be broken to one of its subgroups.
The cases with
$SU(5)\times U(1)$ (flipped $SU(5)$) \cite{revamp},
$SO(6)\times SO(4)$ (Pati--Salam) \cite{alr},
$SU(3)\times SU(2)\times U(1)^2$ (Standard--like) \cite{fny,eu}
and $SU(3)\times SU(2)^2\times U(1)$ (left--right symmetric) \cite{lrs}
were shown to produce quasi--realistic examples.
The Pati--Salam free fermionic heterotic--string models utilise only periodic and
anti--periodic boundary conditions, whereas the other cases necessarily use also
fractional boundary conditions. The Pati--Salam case \cite{ps}
therefore represents the
simplest extension of the classification program
of \cite{fkr} to quasi--realistic models.

\section{Pati--Salam Heterotic--String Models}\label{analysis}

In the free fermionic formulation the 4-dimensional heterotic string, in the
light-cone gauge, is described
by $20$ left moving  and $44$ right moving real fermions.
A large number of models can be constructed by choosing
different phases picked up by   fermions ($f_A, A=1,\dots,44$) when transported
along the torus non-contractible loops.
Each model corresponds to a particular choice of fermion phases consistent with
modular invariance
that can be generated by a set of  basis vectors $v_i,i=1,\dots,n$
$$v_i=\left\{\alpha_i(f_1),\alpha_i(f_{2}),\alpha_i(f_{3}))\dots\right\}$$
describing the transformation  properties of each fermion
\begin{equation}
f_A\to -e^{i\pi\alpha_i(f_A)}\ f_A, \ , A=1,\dots,44
\end{equation}
The basis vectors span a space $\Xi$ which consists of $2^N$ sectors that give
rise to the string spectrum. Each sector is given by
\begin{equation}
\xi = \sum N_i v_i,\ \  N_i =0,1
\end{equation}
The spectrum is truncated by a generalized GSO projection whose action on a
string state  $|S>$ is
\begin{equation}\label{eq:gso}
e^{i\pi v_i\cdot F_S} |S> = \delta_{S}\ \cc{S}{v_i} |S>,
\end{equation}
where $F_S$ is the fermion number operator and $\delta_{S}=\pm1$ is the
space--time spin statistics index.
Different sets of projection coefficients $\cc{S}{v_i}=\pm1$ consistent with
modular invariance give
rise to different models. Summarizing: a model can be defined uniquely by a set
of basis vectors $v_i,i=1,\dots,n$
and a set of $2^{N(N-1)/2}$ independent projections coefficients
$\cc{v_i}{v_j}, i>j$.

The free fermions in the light-cone gauge in the usual notation are:
$\psi^\mu, \chi^i,y^i, \omega^i, i=1,\dots,6$ (left-movers) and
$\bar{y}^i,\bar{\omega}^i, i=1,\dots,6$,
$\psi^A, A=1,\dots,5$, $\bar{\eta}^B, B=1,2,3$, $\bar{\phi}^\alpha,
\alpha=1,\ldots,8$ (right-movers).
The class of models we investigate, is generated by a set of thirteen basis vectors
$
B=\{v_1,v_2,\dots,v_{13}\},
$
where
\begin{eqnarray}
v_1=1&=&\{\psi^\mu,\
\chi^{1,\dots,6},y^{1,\dots,6}, \omega^{1,\dots,6}| \nonumber\\
& & ~~~\bar{y}^{1,\dots,6},\bar{\omega}^{1,\dots,6},
\bar{\eta}^{1,2,3},
\bar{\psi}^{1,\dots,5},\bar{\phi}^{1,\dots,8}\},\nonumber\\
v_2=S&=&\{\psi^\mu,\chi^{1,\dots,6}\},\nonumber\\
v_{2+i}=e_i&=&\{y^{i},\omega^{i}|\bar{y}^i,\bar{\omega}^i\}, \
i=1,\dots,6,\nonumber\\
v_{9}=b_1&=&\{\chi^{34},\chi^{56},y^{34},y^{56}|\bar{y}^{34},
\bar{y}^{56},\bar{\eta}^1,\bar{\psi}^{1,\dots,5}\},\label{basis}\\
v_{10}=b_2&=&\{\chi^{12},\chi^{56},y^{12},y^{56}|\bar{y}^{12},
\bar{y}^{56},\bar{\eta}^2,\bar{\psi}^{1,\dots,5}\},\nonumber\\
v_{11}=z_1&=&\{\bar{\phi}^{1,\dots,4}\},\nonumber\\
v_{12}=z_2&=&\{\bar{\phi}^{5,\dots,8}\},\nonumber\\
v_{13}=\alpha &=& \{\bar{\psi}^{4,5},\bar{\phi}^{1,2}\}.\nonumber
\end{eqnarray}
The first twelve vectors in this set are identical to those used in \cite{fknr,fkr}.
$v_{13}$ is the additional new vector that breaks the $SO(10)$ GUT symmetry to
$SO(6)\times SO(4)$. The second ingredient that is needed to define the string vacuum
are the GGSO projection coefficients that appear in the one--loop partition function,
$\cc{v_i}{v_j}$, spanning a $13\times 13$ matrix. Only the elements with $i>j$ are
independent, and the others are fixed by modular invariance.
A priori there are therefore 78 independent coefficients corresponding
to $2^{78}$ distinct string vacua. Eleven coefficients
are fixed by requiring that the models possess $N=1$ supersymmetry.
Additionally, we impose the
condition that the only space--time vector bosons that remain in the spectrum are those
that arise from the untwisted sector.
This restricts further the number of phases,
leaving a total of 51 independent GGSO phases.
The gauge group in these models is therefore:
\beqn
{\rm observable} ~: &~~~~~~~~SO(6)\times SO(4) \times U(1)^3 \nonumber\\
{\rm hidden}     ~: &~~SO(4)^2\times SO(8)~~~~             \nonumber
\eeqn

The untwisted matter is common in these models and is composed of three pairs
of vectorial representations of the observable $SO(6)$ symmetry, and $SO(10)$
singlets.
The chiral matter spectrum arises from the twisted sectors.
The chiral spinorial representations of the observable $SO(6) \times SO(4)$
arise from the sectors:
\begin{eqnarray}
B_{\ell_3^1\ell_4^1\ell_5^1\ell_6^1}^1&=&S+b_1+\ell_3^1 e_3+\ell_4^1 e_4 +
\ell_5^1 e_5 + \ell_6^1 e_6 \\
B_{\ell_1^2\ell_2^2\ell_5^2\ell_6^2}^2&=&S+b_2+\ell_1^2 e_1+\ell_2^2 e_2 +
\ell_5^2 e_5 + \ell_6^2 e_6 \label{ss}\\
B_{\ell_1^3\ell_2^3\ell_3^3\ell_4^3}^3&=&
S+b_3+ \ell_1^3 e_1+\ell_2^3 e_2 +\ell_3^3 e_3+ \ell_4^3 e_4
\end{eqnarray}
where $\ell_i^j=0,1$;
$b_3=b_1+b_2+x=1+S+b_1+b_2+\sum_{i=1}^6 e_i+\sum_{n=1}^2 z_n$, and
$x$ is given by the vector $x=\{{\bar\psi}^{1,\cdots,5},{\bar\eta}^{1,2,3}\}$.
These sectors give rise to 16 and $\overline{16}$
representations of $SO(10)$ decomposed under
$SO(6)\times SO(4)\equiv SU(4)\times SU(2)_L\times SU(2)_R$
\beqn
16 = & (4,2,1) + ({\bar4}, 1, 2) \nonumber\\
\overline{16} = & ({\bar4},2,1) + (4,1,2) \nonumber
\eeqn

We note that in these models there are three $SO(4)$
group factors, and there a cyclic symmetry
among them. We could have therefore defined one of the other two $SO(4)$ group as the
observable one, and the other two as the hidden ones. We follow here the convention
that keeps the group generated by the world--sheet fermions ${\bar\psi}^{4,5}$ as
the observable $SO(4)$ and the ones generated by ${\bar\phi}^{1,2}$ and ${\bar\phi}^{3,4}$
as hidden. The models then give rise to multitude of sectors that
produce exotic states with fractional electric charge, given by:
\beq
Q_{em} = {1\over\sqrt{6}}T_{15}+{1\over2}I_{3_L}+{1\over2}I_{3_R}
\eeq
where $T_{15}$ is the diagonal generator of $SU(4)/SU(3)$ and
$I_{3_L}$, $I_{3_R}$
are the diagonal generators of $SU(2)_L$, $SU(2)_R$, respectively.
The models then contain the exotic states in the representations:
\begin{align}
({4},{1},{1})+({\bar{4}},
{1},{1}):&\pm\frac{1}{6} \ \text{exotic coloured particles and
SM singlets}\nn\\
({1},{2},{1}):& \pm\frac{1}{2}\ \text{leptons}\nn\\
 ({1},{1},{2}):& \pm\frac{1}{2}\  \text{SM singlets}\nn
\end{align}

We now enumerate the sectors that give rise to exotic states.
The states corresponding to the representations
$({4},{2},{1})$,
$({4},{1},{2})$,
$(\bar{{4}},{2},{1})$,
$(\bar{{4}},{1},{2})$ where
${4}$ and $\bar{{4}}$ are spinorial (anti--spinorial)
representations of the observable SO(6), and the $2$ are
doublet representations
of the hidden $SU(2)\times SU(2) = SO(4)_1$, arise from the following sectors:

\begin{eqnarray}
B_{pqrs}^{(4)}&=& S + b_1 + b_2 + \beta + p e_1+ q e_2 + r e_3 + s e_4 \nonumber\\
B_{pqrs}^{(5)}&=& S + b_1 + b_3 + \beta + p e_1+ q e_2 + r e_5 + s e_6  \nonumber\\
B_{pqrs}^{(6)}&=& S + b_2 + b_3 + \beta + p e_3+ q e_4 + r e_5 + s e_6~,
\end{eqnarray}

where 
$\beta=\alpha+x
\equiv\{{\bar\psi}^{1,2,3}, {\bar\eta}^{1,2,3},{\bar\phi}^{1,2}\}$.
Similar states arise from the sectors $B_{pqrs}^{(4,5,6)}+ z_1$
and correspond to the representations
$({4},{2},{1})$,
$({4},{1},{2})$,
$(\bar{{4}},{2},{1})$,
$(\bar{{4}},{1},{2})$ of
$SO(6)_{obs} \times SO(4)_2$.

The states corresponding to the representations
$(({2},{1}),({2},{1}))$,
$(({2},{1}),({1},{2}))$,
$(({1},{2}),({1},{2}))$ and
$(({1},{2}),({2},{1}))$ transforming under
 $SU(2)_L\times SU(2)_R \times SO(4)_1$ arise from the sectors:

\begin{eqnarray}
B_{pqrs}^{(7)}&=& B_{pqrs}^{(4)} + x \,\,=\,\,
                  S + b_3 + \beta + p e_1+ q e_2 + r e_3 + s e_4 \nonumber\\
B_{pqrs}^{(8)}&=& B_{pqrs}^{(5)} + x \,\,=\,\,
                  S + b_2 + \beta + p e_1+ q e_2 + r e_5 + s e_6  \nonumber\\
B_{pqrs}^{(9)}&=& B_{pqrs}^{(6)} + x \,\,=\,\,
                  S + b_1 + \beta + p e_3+ q e_4 + r e_5 + s e_6
\end{eqnarray}

The remaining sectors give rise to states that transform as representations of the hidden
gauge group, and are singlets under the observable $SO(10)$ GUT symmetry. These states
are therefore hidden matter states that arise in the string model, but are not
exotic with respect to electric charge.
The following 48 sectors produce the representations
$(({2},{1}),({2},{1}))$ of
$SU(2)^4 = SO(4)_1 \times SO(4)_2$:
\begin{eqnarray}
B_{pqrs}^{(10)} &=& B_{pqrs}^{(1)} + x + z_1 \,\,=\,\,
                S + b_2 + b_3 + p e_3+ q e_4 + r e_5 + s e_6 + z_1 \nonumber\\
B_{pqrs}^{(11)} &=& B_{pqrs}^{(2)} + x + z_1 \,\,=\,\,
                S + b_1 + b_3 + p e_1+ q e_2 + r e_5 + s e_6 + z_1 \nonumber\\
B_{pqrs}^{(12)} &=& B_{pqrs}^{(3)} + x + z_1 \,\,=\,\,
                S + b_1 + b_2 + p e_1+ q e_2 + r e_3 + s e_4 + z_1
\end{eqnarray}
There are 48 sectors producing spinorial ${8}$
and anti--spinorial $\bar{{8}}$ representations of the hidden
$SO(8)$ gauge group:
\begin{eqnarray}
B_{pqrs}^{(13)} &=& B_{pqrs}^{(1)} + x + z_2 \,\,=\,\,
                S + b_2 + b_3 + p e_3+ q e_4 + r e_5 + s e_6 + z_2 \nonumber\\
B_{pqrs}^{(14)} &=& B_{pqrs}^{(2)} + x + z_2 \,\,=\,\,
                S + b_1 + b_3 + p e_1+ q e_2 + r e_5 + s e_6 + z_2 \nonumber\\
B_{pqrs}^{(15)} &=& B_{pqrs}^{(3)} + x + z_2 \,\,=\,\,
                S + b_1 + b_2 + p e_1+ q e_2 + r e_3 + s e_4 + z_2
\end{eqnarray}

States that transform in vectorial representations are obtained from sectors that
contain four periodic world--sheet right--moving complex fermions. Massless states
are obtained in such sectors by acting on the vacuum with a Neveu--Schwarz
right--moving fermionic oscillator. Vectorial representations arise from the
sectors:
\begin{eqnarray}
B_{pqrs}^{(1)}+x&=& S + b_1 + x + p e_3+ q e_4 + r e_5 + s e_6 \nonumber\\
B_{pqrs}^{(2)}+x&=& S + b_2 + x + p e_1+ q e_2 + r e_5 + s e_6  \nonumber\\
B_{pqrs}^{(3)}+x&=& S + b_3 + x + p e_1+ q e_2 + r e_3 + s e_4
\end{eqnarray}
and produce the following representations:

\begin{itemize}
\item $\{\bar{\psi}^{123}\}|R>_{pqrs}^{(i)}$, $i = 1,2,3$, where
$|R>_{pqrs}^{(i)}$ is the degenerated Ramond vacuum of the $B_{pqrs}^{(i)}$ sector.
These states transform as a vectorial representation of SO(6).
\item $\{\bar{\psi}^{45}\}|R>_{pqrs}^{(i)}$, $i = 1,2,3$, where
$|R>_{pqrs}^{(i)}$ is the degenerated Ramond vacuum of the $B_{pqrs}^{(i)}$ sector.
These states transform as a vectorial representation of SO(4).
\item $\{\bar{\phi}^{12}\}|R>_{pqrs}^{(i)}$, $i = 1,2,3$.
These states transform as a vectorial representation of SO(4).
\item $\{\bar{\phi}^{34}\}|R>_{pqrs}^{(i)}$, $i = 1,2,3$.
These states transform as a vectorial representation of SO(4).
\item $\{\bar{\phi}^{5..8}\}|R>_{pqrs}^{(i)}$, $i = 1,2,3$.
These states transform as a vectorial representation of SO(8).
\item the remaining states in those sectors transform as singlets
of the non--Abelian group factors.
\end{itemize}

\section{Exophobic String Models}\label{exophobia}

Following the methodology developed in \cite{fkr} we can
write down analytic expressions for the GGSO projections
on the states arising from all the sectors listed above.
These formulas are inputted into a computer program which is used to
scan the space of string vacua generated by random generation
of the one--loop GGSO projection coefficients.
The number of possible configurations is $2^{51}\sim 10^{15}$, which is
too large for a complete classification. For this reason
a random generation algorithm is utilised, and a model with the
desired phenomenological criteria is fished from the sample generated.

The observable sector of a
heterotic--string Pati--Salam model is characterized by 9 integers
$\left(n_g,k_L,k_R,n_6,n_h,n_4,n_{\bar{4}},n_{2L},n_{2R}\right)$, where
\begin{align}
&n_{4L}-n_{\bar{4}L}=n_{\bar{4}R}-n_{4R}=n_g=\text{\# of generations}\nn\\
&n_{\bar{4}L}=k_L=\text{\# of non chiral left pairs}\nn\\
&n_{4R}=k_R=\text{\# of non chiral right pairs}\nn\\
&n_6=\text{\# of } ({6},{1},{1})\nn\\
&n_h=\text{\# of } ({1},{2},{2})\nn\\
&n_4=\text{\# of } ({4},{1},{1})\text{ (exotic)}\nn\\
&n_{\bar{4}}=\text{\# of } ({\bar{4}},{1},{1})\text{ (exotic)}\nn\\
&n_{2L}=\text{\# of } ({1},{2},{1})\ \text{ (exotic)}\nn\\
&n_{2R}=\text{\# of } ({1},{1},{2}) \text{ (exotic)}\nn
\end{align}
Following the methodology developed in \cite{fkr} we derived analytic formulas
for all these quantities, and will reported elsewhere.
The spectrum of a viable Pati--Salam heterotic string model should
have $n_g=3$,
\begin{align}
&n_g=3~~~\text{three light chiral of generations}\nn\\
&k_L\ge0~~~\text{heavy mass can be generated for non chiral pairs}\nn\\
&k_R\ge1~~~\text{at least one Higgs pair to break the PS symmetry}\nn\\
&n_6\ge1~~~\text{at least one required for missing partner mechanism }\nn\\
&n_h\ge1~~~\text{at least one light Higgs bi--doublet}\nn\\
&n_4=n_{\bar{4}}\ge0~~~\text{heavy mass can be generated for vector--like exotics}\nn\\
&n_{2L}=0\text{mod}2~~~\text{heavy mass can be generated for vector--like exotics}\nn\\
&n_{2R}=0\text{mod}2~~~\text{heavy mass can be generated for vector--like exotics}\nn
\end{align}

A minimal model which is free of exotics has
$k_L=0$, $k_R=1$,
$n_6=1$, $n_h=3$, $n_4=n_{\bar{4}}=0$, $n_{2L}=0$ and $n_{2R}=0$.
The model given by the following GGSO coefficients matrix :

\beq [v_i|v_j] = e^{i\pi (v_i|v_j)} \eeq

\beq \label{BigMatrix}  (v_i|v_j)\ \ =\ \ \bordermatrix{
        & 1& S&e_1&e_2&e_3&e_4&e_5&e_6&b_1&b_2&z_1&z_2&\alpha\cr
 1  & 1& 1& 1& 1& 1& 1& 1& 1& 1& 1& 1& 1 &0\cr
S  & 1& 1& 1& 1& 1& 1& 1& 1& 1& 1& 1& 1& 1\cr
e_1& 1& 1& 0& 1& 0& 0& 0& 0& 0& 0& 0& 1& 1\cr
e_2& 1& 1& 1& 0& 1& 1& 1& 0& 1& 0& 1& 1& 0\cr
e_3& 1& 1& 0& 1& 0& 1& 1& 0& 0& 1& 0& 0& 0\cr
e_4& 1& 1& 0& 1& 1& 0& 1& 0& 0& 1& 0& 0& 1\cr
e_5& 1& 1& 0& 1& 1& 1& 0& 1& 1& 0& 0& 0& 0\cr
e_6& 1& 1& 0& 0& 0& 0& 1& 0& 0& 0& 0& 1& 1\cr
b_1& 1& 0& 0& 1& 0& 0& 1& 0& 1& 1& 0& 0& 1\cr
b_2& 1& 0& 0& 0& 1& 1& 0& 0& 1& 1& 1& 1& 0\cr
z_1& 1& 1& 0& 1& 0& 0& 0& 0& 0& 1& 1& 1& 1\cr
z_2& 1& 1& 1& 1& 0& 0& 0& 1& 0& 1& 1& 1& 0\cr
\alpha& 0& 1& 1& 0& 0& 1& 0& 1& 0& 1& 0& 0& 0\cr
  }
\eeq
The twisted massless states generated in the string vacuum
of eq. (\ref{BigMatrix}) produces the desired spectrum. Namely, it contains
three chiral generations; one pair of heavy Higgs states to break the Pati--Salam
symmetry along flat direction; one light Higgs bi-doublet to break the electroweak
symmetry and generate fermion masses; one vector sextet of $SO(6)$ needed
for the missing partner mechanism; it is completely free of massless exotic fractionally charged
states. States with fractional electric charge necessarily exist in the massive spectrum of the
string model, and it easy to see that indeed they do. All we need is to show that the
GSO projection which projects
them in a given sector is reversed when additional
Neveu--Schwarz oscillators act on the vacuum.
Additionally the model contains three pairs of untwisted $SO(6)$ sextets.
These can obtain string scale mass along flat directions.
The full massless spectrum of the model is shown in tables
\ref{tablea} and \ref{tableb}.

\begin{table}
\noindent
{\small
\openup\jot
\begin{tabular}{|l|l|c|c|c|c|}
\hline
sector&field&$SU(4)\times{SU(2)}_L\times{SU(2)}_R$&${U(1)}_1$&${U(1)}_2$&${U(1)}_3$\\
\hline
\hline
$S$&$D_1$&$(6,1,1)$&$+1$&$\hphantom{+}0$&$\hphantom{+}0$\\
&$D_2$&$(6,1,1)$&$\hphantom{+}0$&$+1$&$\hphantom{+}0$\\
&$D_3$&$(6,1,1)$&$\hphantom{+}0$&$\hphantom{+}0$&$+1$\\
&$\bar{D}_1$&$(6,1,1)$&$-1$&$\hphantom{+}0$&$\hphantom{+}0$\\
&$\bar{D}_2$&$(6,1,1)$&$\hphantom{+}0$&$-1$&$\hphantom{+}0$\\
&$\bar{D}_3$&$(6,1,1)$&$\hphantom{+}0$&$\hphantom{+}0$&$-1$\\
&$\Phi_{12}$&$(1,1,1)$&$+1$&$+1$&$\hphantom{+}0$\\
&$\Phi_{12}^{-}$&$(1,1,1)$&$+1$&$-1$&$\hphantom{+}0$\\
&$\bar{\Phi}_{12}$&$(1,1,1)$&$-1$&$-1$&$\hphantom{+}0$\\
&$\bar{\Phi}_{12}^{-}$&$(1,1,1)$&$-1$&$+1$&$\hphantom{+}0$\\
&$\Phi_{13}$&$(1,1,1)$&$+1$&$\hphantom{+}0$&$+1$\\
&$\Phi_{13}^-$&$(1,1,1)$&$+1$&$\hphantom{+}0$&$-1$\\
&$\bar{\Phi}_{13}$&$(1,1,1)$&$-1$&$\hphantom{+}0$&$-1$\\
&$\bar{\Phi}_{13}^-$&$(1,1,1)$&$-1$&$\hphantom{+}0$&$+1$\\
&$\Phi_i,i=1,\dots,6$&$(1,1,1)$&$\hphantom{+}0$&$\hphantom{+}0$&$\hphantom{+}0$\\
&$\Phi_{23}$&$(1,1,1)$&$\hphantom{+}0$&$+1$&$+1$\\
&$\Phi_{23}^-$&$(1,1,1)$&$\hphantom{+}0$&$+1$&$-1$\\
&$\bar{\Phi}_{23}$&$(1,1,1)$&$\hphantom{+}0$&$-1$&$-1$\\
&$\bar{\Phi}_{23}^-$&$(1,1,1)$&$\hphantom{+}0$&$-1$&$+1$\\
\hline
$S+b_1+e_5$&${F}_{1L}$&$({4},2,1)$&$\hphantom{+}\frac{1}{2}$&$\hphantom{+}0$&$\hphantom{+}0$\\
\hline
$S+b_1+e_4$&$\bar{F}_{1R}$&$(\bar{4},1,2)$&$-\frac{1}{2}$&$\hphantom{+}0$&$\hphantom{+}0$\\
\hline
$S+b_1+e_3$&${F}_{2L}$&$({4},2,1)$&$-\frac{1}{2}$&$\hphantom{+}0$&$\hphantom{+}0$\\
\hline
$S+b_1+e_3+e_4+e_5$&$\bar{F}_{2R}$&$(\bar{4},1,2)$&$\hphantom{+}\frac{1}{2}$&$\hphantom{+}0$&$\hphantom{+}0$\\
\hline
$S+b_2+e_2$&${F}_{1R}$&$({4},1,2)$&$\hphantom{+}0$&$-\frac{1}{2}$&$\hphantom{+}0$\\
\hline
$S+b_2+e_1+e_2+e_6$&$\bar{F}_{3R}$&$(\bar{4},1,2)$&$\hphantom{+}0$&$\hphantom{-}\frac{1}{2}$&$\hphantom{+}0$\\
\hline
$S+b_3+e_4$&${F}_{3L}$&$({4},2,1)$&$\hphantom{+}0$&$\hphantom{+}0$&$-\frac{1}{2}$\\
\hline
$S+b_3+e_3$&$\bar{F}_{4R}$&$(\bar{4},1,2)$&$\hphantom{+}0$&$\hphantom{+}0$&$-\frac{1}{2}$\\
\hline
$S+b_1+b_2+e_2$&$h_1$&$(1,2,2)$&$-\frac{1}{2}$&$-\frac{1}{2}$&$\hphantom{+}0$\\
\hline
$S+b_1+b_3+e_1+e_6$&$h_2$&$(1,2,2)$&$-\frac{1}{2}$&$\hphantom{+}0$&$-\frac{1}{2}$\\
\hline
$S+b_1+b_3$&$h_3$&$(1,2,2)$&$\hphantom{+}\frac{1}{2}$&$\hphantom{+}0$&$\hphantom{+}\frac{1}{2}$\\
\hline
$S+b_1+b_2+e_2+e_3+e_4$&$D_4$&$(6,1,1)$&$-\frac{1}{2}$&$-\frac{1}{2}$&$\hphantom{+}0$\\
&$\zeta_a, a=1,2$&$(1,1,1)$&$\hphantom{+}\frac{1}{2}$&$-\frac{1}{2}$&$\hphantom{+}0$\\
&$\bar{\zeta}_a, a=1,2$&$(1,1,1)$&$-\frac{1}{2}$&$\hphantom{+}\frac{1}{2}$&$\hphantom{+}0$\\
&$\xi_1$&$(1,1,1)$&$\hphantom{+}\frac{1}{2}$&$\hphantom{+}\frac{1}{2}$&$\hphantom{+}1$\\
&${\xi}_2$&$(1,1,1)$&$\hphantom{+}\frac{1}{2}$&$\hphantom{+}\frac{1}{2}$&$-1$\\
\hline
$S+b_1+b_2+e_2+e_4$&$\zeta_3$&$(1,1,1)$&$\hphantom{+}\frac{1}{2}$&$\hphantom{+}\frac{1}{2}$&$\hphantom{+}0$\\
$$&$\bar{\zeta}_3$&$(1,1,1)$&$-\frac{1}{2}$&$-\frac{1}{2}$&$\hphantom{+}0$\\
\hline
\end{tabular}
}
\caption{\label{tablea}\it
Observable sector spectrum
$SU(4)\times{SU(2)}_L\times{SU(2)}_R\times{U(1)}^3$ quantum numbers.}
\end{table}
\begin{table}
\noindent
{\small
\begin{tabular}{|l|l|c|c|c|c|}
\hline
sector&field&${SU(2)}^4\times{SO(8)}$&${U(1)}_1$&${U(1)}_2$&${U(1)}_3$\\
\hline
$S+b_1+b_2+e_1+e_4$&$H_{12}^1$&$(2,2,1,1,1)$&$-\frac{1}{2}$&$-\frac{1}{2}$&$\hphantom{+}0$\\\hline
$S+b_2+b_3+e_3+e_6$&$H_{12}^2$&$(2,2,1,1,1)$&$\hphantom{+}0$&$-\frac{1}{2}$&$-\frac{1}{2}$\\\hline
$S+b_2+b_3+e_5+e_6$&$H_{12}^3$&$(2,1,2,1,1)$&$\hphantom{+}0$&$-\frac{1}{2}$&$\hphantom{+}\frac{1}{2}$\\\hline
$S+b_1+b_2+z_1+e_4$&$H_{13}^1$&$(2,1,2,1,1)$&$-\frac{1}{2}$&$\hphantom{+}\frac{1}{2}$&$\hphantom{+}0$\\\hline
$S+b_2+b_3+z_1+e_3+e_5$&$H_{13}^2$&$(2,1,2,1,1)$&$\hphantom{+}0$&$\hphantom{+}\frac{1}{2}$&$-\frac{1}{2}$\\\hline
$S+b_2+b_3+z_1$&$H_{13}^3$&$(2,1,2,1,1)$&$\hphantom{+}0$&$-\frac{1}{2}$&$-\frac{1}{2}$\\\hline
$S+b_1+b_2+z_1+e_1+e_2$&$H_{14}^1$&$(2,1,1,2,1)$&$\hphantom{+}\frac{1}{2}$&$\hphantom{+}\frac{1}{2}$&$\hphantom{+}0$\\\hline
$S+b_1+b_3+z_1+e_1$&$H_{14}^2$&$(2,1,1,2,1)$&$-\frac{1}{2}$&$\hphantom{+}0$&$-\frac{1}{2}$\\\hline
$S+b_1+b_3+z_1+e_6$&$H_{14}^3$&$(2,1,1,2,1)$&$\hphantom{+}\frac{1}{2}$&$\hphantom{+}0$&$\hphantom{+}\frac{1}{2}$\\\hline
$S+b_1+b_2+z_1+e_1+e_2+e_3+e_4$&$H_{23}^1$&$(1,2,2,1,1)$&$\hphantom{+}\frac{1}{2}$&$\hphantom{+}\frac{1}{2}$&$\hphantom{+}0$\\\hline
$S+b_1+b_2+z_1+e_3$&$H_{24}^1$&$(1,2,1,2,1)$&$-\frac{1}{2}$&$\hphantom{+}\frac{1}{2}$&$\hphantom{+}0$\\\hline
$S+b_1+b_3+z_1+e_1+e_2+e_5+e_6$&$H_{24}^2$&$(1,2,2,1,1)$&$-\frac{1}{2}$&$\hphantom{+}0$&$-\frac{1}{2}$\\\hline
$S+b_1+b_3+z_1+e_2+e_5$&$H_{24}^3$&$(1,2,2,1,1)$&$\hphantom{+}\frac{1}{2}$&$\hphantom{+}0$&$\hphantom{+}\frac{1}{2}$\\\hline
$S+b_2+b_3+z_1+e_3+e_4$&$H_{24}^4$&$(1,2,1,2,1)$&$\hphantom{+}0$&$-\frac{1}{2}$&$-\frac{1}{2}$\\\hline
$S+b_2+b_3+z_1+e_4+e_5$&$H_{24}^5$&$(1,2,1,2,1)$&$\hphantom{+}0$&$\hphantom{+}\frac{1}{2}$&$-\frac{1}{2}$\\\hline
$S+b_1+b_2+e_1+e_3$&$H_{34}^1$&$(1,1,2,2,1)$&$-\frac{1}{2}$&$\hphantom{+}\frac{1}{2}$&$\hphantom{+}0$\\\hline
$S+b_1+b_2+e_1+e_2+e_5$&$H_{34}^2$&$(1,1,2,2,1)$&$\hphantom{+}\frac{1}{2}$&$\hphantom{+}0$&$-\frac{1}{2}$\\\hline
$S+b_2+b_3+e_2+e_5+e_6$&$H_{34}^3$&$(1,1,2,2,1)$&$-\frac{1}{2}$&$\hphantom{+}0$&$-\frac{1}{2}$\\\hline
$S+b_2+b_3+e_3+e_4+e_5+e_6$&$H_{34}^4$&$(1,1,2,2,1)$&$\hphantom{+}0$&$-\frac{1}{2}$&$\hphantom{+}\frac{1}{2}$\\\hline
$S+b_2+b_3+e_4+e_6$&$H_{34}^5$&$(1,1,2,2,1)$&$\hphantom{+}0$&$-\frac{1}{2}$&$-\frac{1}{2}$\\\hline
$S+b_1+b_3+e_6$&$Z_1$&$(1,1,1,1,8_v)$&$\hphantom{+}\frac{1}{2}$&$\hphantom{+}0$&$-\frac{1}{2}$\\\hline
$S+b_1+b_3+z_2+e_1+e_2+e_5$&$Z_2$&$(1,1,1,1,8_s)$&$-\frac{1}{2}$&$\hphantom{+}0$&$-\frac{1}{2}$\\\hline
$S+b_1+b_3+z_2+e_1$&$Z_3$&$(1,1,1,1,8_v)$&$-\frac{1}{2}$&$\hphantom{+}0$&$\hphantom{+}\frac{1}{2}$\\\hline
$S+b_1+b_3+z_2+e_2+e_5+e_6$&$Z_4$&$(1,1,1,1,8_s)$&$-\frac{1}{2}$&$\hphantom{+}0$&$-\frac{1}{2}$\\\hline
$S+b_1+b_2+e_1+e_2$&$Z_5$&$(1,1,1,1,8_s)$&$\hphantom{+}\frac{1}{2}$&$-\frac{1}{2}$&$\hphantom{+}0$\\
\hline
\end{tabular}
}
\caption{\label{tableb}\it Hidden sector spectrum and ${SU(2)}^4\times{SO(8)}\times{U(1)}^3$ quantum numbers.}
\end{table}

\begin{figure}[!ht]
\centering
\includegraphics[width=10cm]{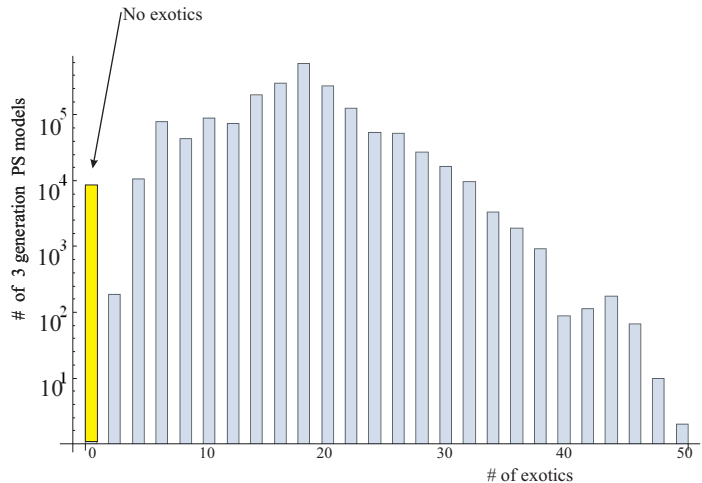}
\caption{\label{exoticsin3genmodels}
\it Number of three generation models versus the total number of exotics.}
\end{figure}

To explore the abundance of string vacua that do not contain massless exotics
we perform a statistical sampling\footnote{We note that analysis of large
sets of string vacua has also been carried
out by other groups \cite{statistical}.}
in a space of $5\times 10^{9}$ models out of the
total of $2^{51}\sim10^{15}$. In figure \ref{exoticsin3genmodels}
we display the
total number of exotics that appear in viable three generation models.
As can be seen
from the figure, in the space of models sampled there are of the order
of $10^4$ models
that are completely free of exotic states.
Having established a quasi--realistic spectrum, the next stage is to
analyze the Yukawa couplings in the models. The abundance of exotic free
three generation models suggests that models with viable Yukawa couplings
and fermion mass spectrum do exist in this space of string vacua.

\section{Conclusions}\label{conclude}

In this paper we demonstrated the existence of string vacua that are
completely free of massless exotic fields. The need to break the
non--Abelian GUT symmetries in heterotic--string models
by Wilson lines, while preserving the GUT embedding of the
weak--hypercharge and the GUT prediction
$\sin^2\theta_w(M_{\rm GUT})=3/8$, necessarily implies that the models
contain states with fractional electric charge. Such states are
severely restricted by observations, and must be confined or
sufficiently massive and diluted. In this paper we constructed
the first quasi--realistic heterotic--string models in which the
exotic states do not appear in the massless spectrum, and only exist,
as they must, in the massive spectrum. The string models that we
constructed are three generation models in which the $SO(10)$ GUT symmetry
is broken to the Pati--Salam subgroup, and similar analysis can be performed
in the case of the other $SO(10)$ subgroups. Our PS heterotic--string
models contain adequate Higgs representations to break the GUT and electroweak
symmetry, as well as colour Higgs triplets that can be used for the missing
partner mechanism. By statistically sampling the space of Pati--Salam
free fermionic vacua we demonstrated the abundance of three generation models
that are completely free of massless exotics, rendering it plausible that
obtaining realistic Yukawa couplings may be possible in this space of models.

\section{Acknowledgements}

AEF and JR would like to thank the Ecole Normal Superier, and BA, CK and JR
would like to thank the University of Liverpool, for hospitality.
AEF is supported in part by STFC under contract PP/D000416/1.
CK is supported in part by the EU under the contracts
MRTN-CT-2004-005104, MRTN-CT-2004-512194,
MRTN-CT-2004-503369, MEXT-CT-2003-509661, CNRS PICS 2530, 3059 and
3747, ANR (CNRS-USAR) contract  05-BLAN-0079-01.
JR work is supported in part by the EU under contracts
MRTN--CT--2006--035863--1 and  MRTN--CT--2004--503369.



\bigskip
\medskip

\bibliographystyle{unsrt}

\end{document}